# Insight-HXMT science operations

S. M. Jia, X. Ma, Y. Huang, W. Z. Zhang, G. Ou, et al.

# Insight-HXMT Science Operations

S.M. Jia*[a], X. Ma[a], Y. Huang[a], W.Z. Zhang[b], G. Ou[a],
L.M. Song[a], J.L. Qu[a], S. Zhang[a], L. Chen[b]

[a]Key Laboratory of Particle Astrophysics, Institute of High Energy Physics, Beijing 100049, China
[b]Beijing Normal University, Beijing 100875, China

## ABSTRACT

The Hard X-ray Modulation Telescope (Insight-HXMT) was successfully launched on June 15th, 2017. It performs broad band X-ray scan survey of the Galactic Plane to detect new black holes and other objects in active states. It also observes X-ray binaries to study their X-ray variabilities. Here we will introduce the Science Operations of Insight-HXMT, which is responsible for collecting and evaluating observation proposals, scheduling observations, and monitoring the working status of the payloads

## 1. INTRODUCTION

The Hard X-ray Modulation Telescope [1] (Insight-HXMT) is the first X-ray astronomy satellite in China, which was successfully launched to an orbit of 550 km on June 15th, 2017. Insight-HXMT contains three instruments: High Energy X-ray Telescope (HE, 20-250 keV), Medium Energy X-ray Telescope (ME, 5-30 keV), and Low Energy X-ray Telescope (LE, 1-15keV). Its core sciences are: (1) to perform broad band X-ray scan survey of the Galactic Plane to detect new black holes and other objects in active states, (2) to observe X-ray binaries to study their X-ray variabilities, (3) observe GRB at an explored band from a few hundred keV to a few MeV.

Insight-HXMT Ground Segment (HGS) is mainly in charge of the payload operation, the observation scheduling, the data receiving, data products, and the user support. HGS is an important guarantee for the realization of the scientific objective of Insight-HXMT, which consists of two parts: the Mission Operation Ground Segment (HMOGS) and the Science Ground Segment (HSGS), see Figure 1.

HMOGS, located in National Space Science center, CAS, includes the Mission Operation Center (MOC), the Data Receiving Station (DRS), and the Space Science Data Center (SSDC). MOC is responsible for monitoring the telescope working status and planning on receiving the data. DRS will receive Insight-HXMT data according to the data receiving plan about 4-5 times each day using the three antennas in Beijing, Sanya, and Kashgar. SSDC is in charge of the data pre-processing and the long-term data archiving.

HGS, located in the Institute of High Energy Physics, CAS, includes the Science Support Center (HSSC), the Science User Center (HSUC), the Science Operation Center (HSOC), and the Science Data Center (HSDC). HSSC formulates the observation policy to guide the observation plan. HSUC collects the proposals and also provides the data release and user support. HSOC is responsible for the observation scheduling and the payload monitoring. The main task of HSDC is to generate the standard data products and the calibration database.

HSSC plans the annual observations and publishes the Announcement of Opportunity (AO). HSUC collects the proposals from scientists. Then the technological and scientific evaluations of the proposals will be carried out by HSOC and HSSC. The approved proposals will become the main scientific task of Insight-HXMT. According to the status of Insight-HXMT and the properties of the observation objects, HSOC will formulate the long-term, medium-term, and short-term observation plans and send them to MOC. MOC will determine the corresponding data reception plans and generate the payload commands. Finally, the satellite tracking telemetry and command (TT&C) system will send the commands to Insight-HXMT.

*jiasm@ihep.ac.cn; phone 86 10 88235850; fax 86 10 88235850

Insight-HXMT downlink data will be received by DRS and then forwarded to MOC and SSDC. MOC monitors the working status of the telescope and generates the monitoring data. Then HSOC will analyze the performance variations of the payloads with the monitoring data. On the other hand, SSDC will generate the primary data products from the receiving data and send them the HSDC. HSDC will generate the advanced data products and update the calibration database and the data analysis software when necessary. Then, HSUC and SSDC will publish the data products regularly and also provide the data analysis software and technical supports on the website.

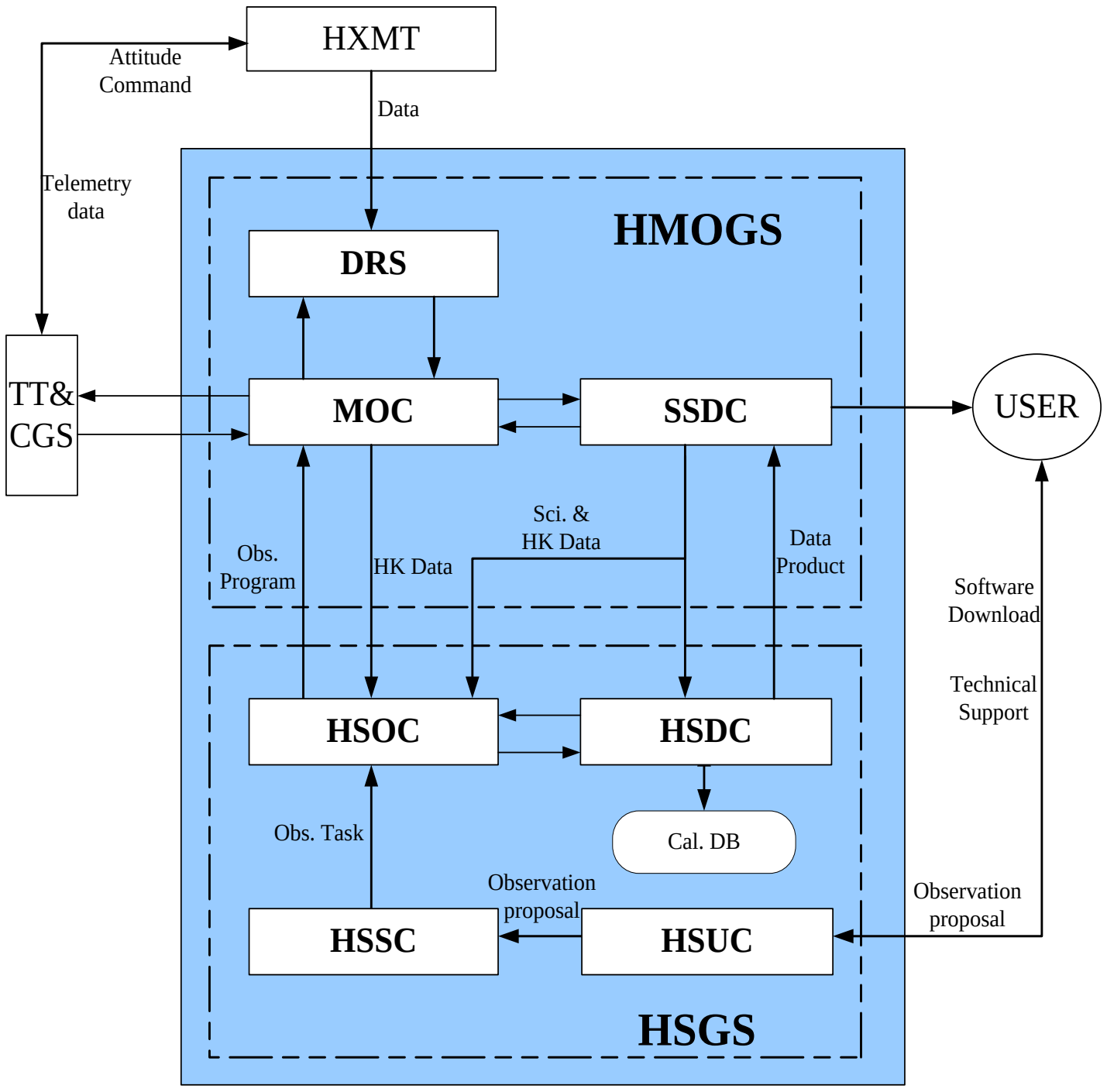

Figure 1. Overview of Insight-HXMT HGS and its workflow.

In this paper, we will focus on the science operation work of Insight-HXMT run by HSOC, including the proposal collection and evaluation, observation scheduling, payloads monitoring, and the observation summary since it was launched.

## 2. PROPOSALS

### 2.1 Proposal collection

In August, 2016, the first call for proposals was announced, and all the proposals were submitted through the website, http://proposal.ihep.ac.cn. The types of the proposals include: (1) Galactic plane survey, (2) High cadence observation of BH and NS systems, (3) High statistics observation of BH and NS systems, (4) Synergy observation with international telescopes, (5) Multi-wavelength coordinated and follow-up observations, (6) others. Totally, we received 90 proposals including 348 sources, the statistics information of these proposals are listed in Table 1.

Table 1. The statistics of proposals in AO01.

| Type | Number | Ratio |
|---|---|---|
| Galactic plane survey | 5 | 6% |
| High cadence observation of BH and NS systems | 15 | 17% |
| High statistics observation of BH and NS systems | 30 | 33% |
| Synergy observation with international telescopes | 1 | 1% |
| Multi-wavelength coordinated and follow-up observations | 12 | 13% |
| others | 27 | 30% |

### 2.2 Proposal evaluation

Considering the observation efficiency and the scientific goals of the proposals, HSOC held the technical and scientific evaluations. Finally, for AO01, 42 normal sources and 116 ToO sources were selected as Grade A, and the total observation time is about 1 year.

## 3. SCHEDULING

### 3.1 Observation constrains

The planning and scheduling of Insight-HXMT observations is a complex multi-objective optimization problem [2]. HSOC need to consider the constrains of Insight-HXMT and the characteristics of the observation sources.

(1) Thermal Control: Shown as Figure 2, the detectors can't be exposed to the sun directly, and Insight-HXMT designed the sunshade to protect the detectors, so the solar avoidance angle should be larger than 70 °, and the angle of the sun to X-Z plane should be less than 10 °.
(2) Earth occultation: The earth occultation will interrupt the observations, and it will occurs almost every orbit, and each time it will last about 30 minutes.
(3) South Atlantic Anomaly (SAA): In SAA, Insight-HXMT will start the SAA mode, and it will occur roughly 8~9 orbits every day, and each time it will last about 15 minutes.
(4) Moon avoidance angle: In order to avoid the moon appearing in the Field of View (FOV) of Insight-HXMT, the moon avoidance angle should be larger than 6 °.

So, the total observation efficiency of Insight-HXMT is about 50%. Figure 3 shows the average efficiency for all the sky in 1 year.

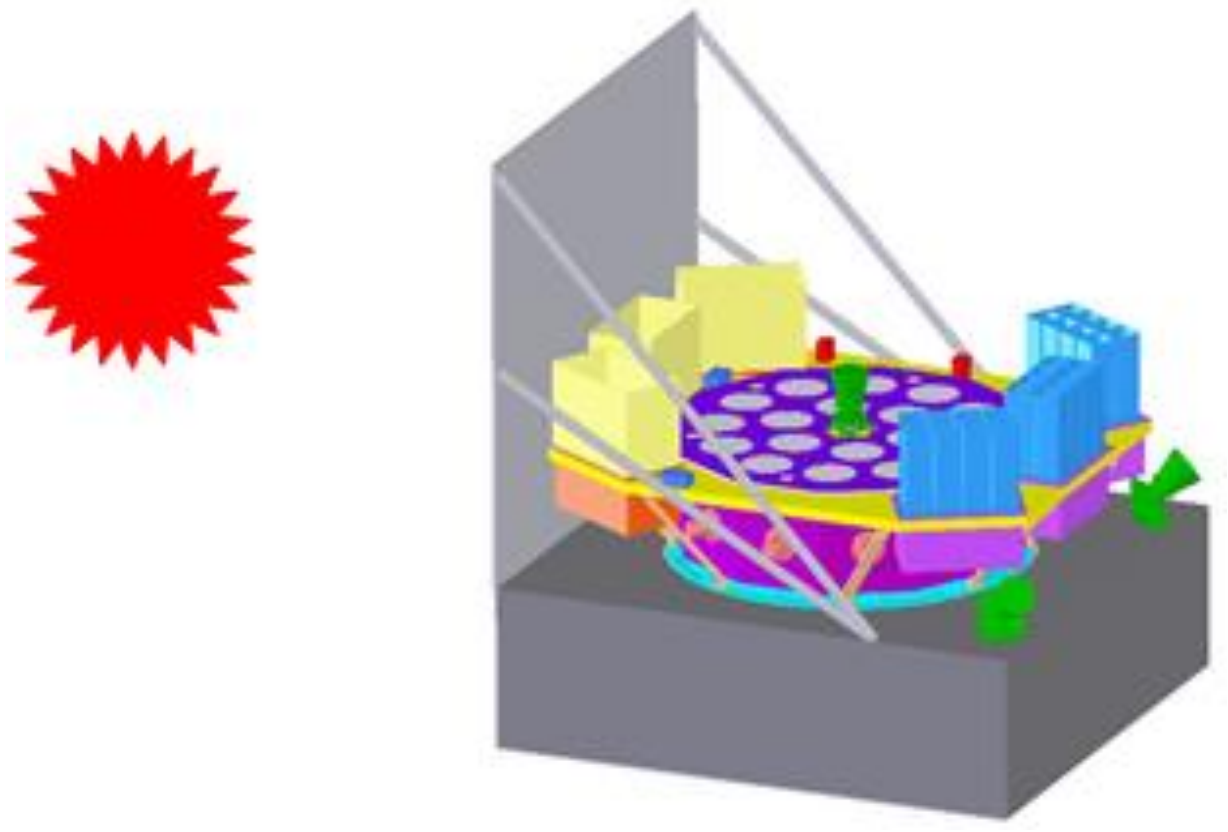

Figure 2. The thermal control of Insight-HXMT.

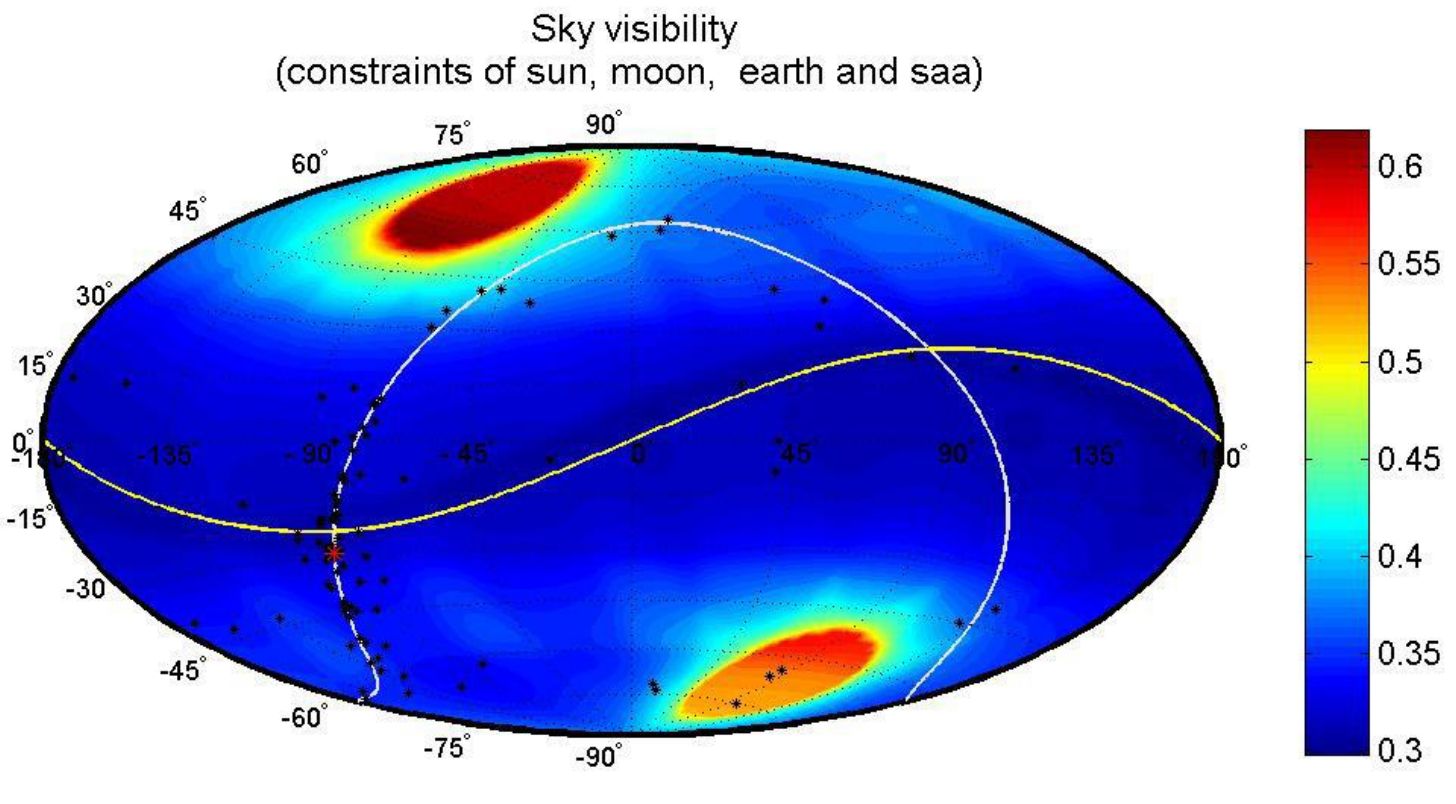

Figure 3. The average efficiency for all the sky in 1 year.

### 3.2 Scheduling

Based on the sources selected from the proposals, HSOC need to schedule the observations. Considering of the science programs and the prediction of the orbit, the plans were divided into three types: the long-term plan, the medium-term plan and the short-term plan, which corresponding to the observational periods of 1 year, 4 weeks and 2 days, with accuracy of up to 4 weeks, 2 days and 1 second, respectively. The short-term plan schedules the observation sources and their sequence, the observation mode and duration, the slew time, the data down-link and so on.

The Insight-HXMT observational plans are available at http://hxmt.org/index.php/plan/splan, and the short-term plan is usually updated every two days.

### 3.3 ToO schedule

Currently, HSOC has two main ways to get the Target of Opportunity (ToO) alert. One is from the astronomy society, and the ToO proposals can be submitted from the website, http://proposal.ihep.ac.cn. The other is from the Galactic Plane survey of Insight-HXMT.

Once HSOC receives a ToO trigger, the ToO workflow should be started. First, HSOC will check the ToO information and carry out the technical evaluation, and this should be finished in 60 mins. If the ToO source is suitable for the observation of Insight-HXMT, HSOC should contact with the Project Scientist who will make the decision if this trigger will be accepted. If the ToO observation is approved, HSOC should make the urgent re-scheduling, and this ToO plan should be finished in 90 mins. Then, HSOC will send the ToO plan to MOC and then the plan will be uplinked to the telescope. So, the response from a ToO trigger to perform a ToO observation, will take about 5 hours. Finally, the ToO observation plan will be updated in the planning website.

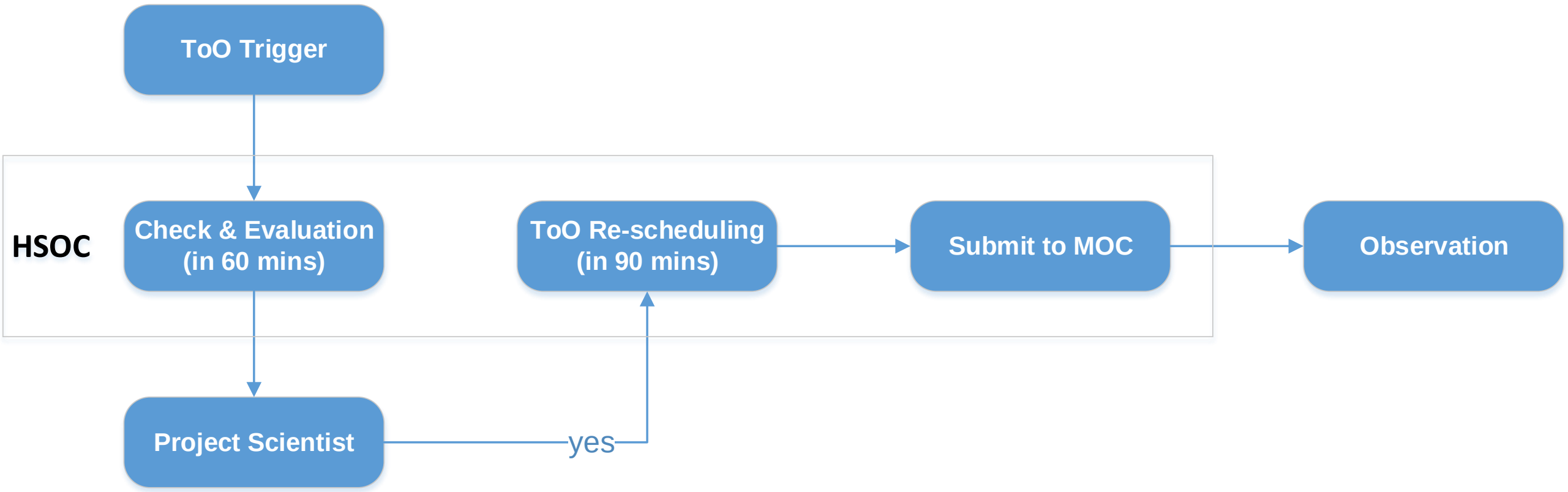

Figure 4. The workflow for ToO observation of Insight-HXMT.

## 4. PAYLOAD MANAGEMENT

### 4.1 Payload monitoring

It was scheduled to receive the observational data of Insight-HXMT for 4-5 times every day, then all the data were sent to MOC. MOC will carry out the near real time monitor for the instrument safety, and then send the data to HSOC.

There are more than 5000 parameters to be monitored for the three instruments of Insight-HXMT. HSOC analyzed the data in detail to monitor: (1) the work status and the health of the payloads with the engineering data, (2) the performance and performance variation trend of the payloads [3] with the science data. Figure 5 shows some monitoring interfaces of Insight-HXMT.

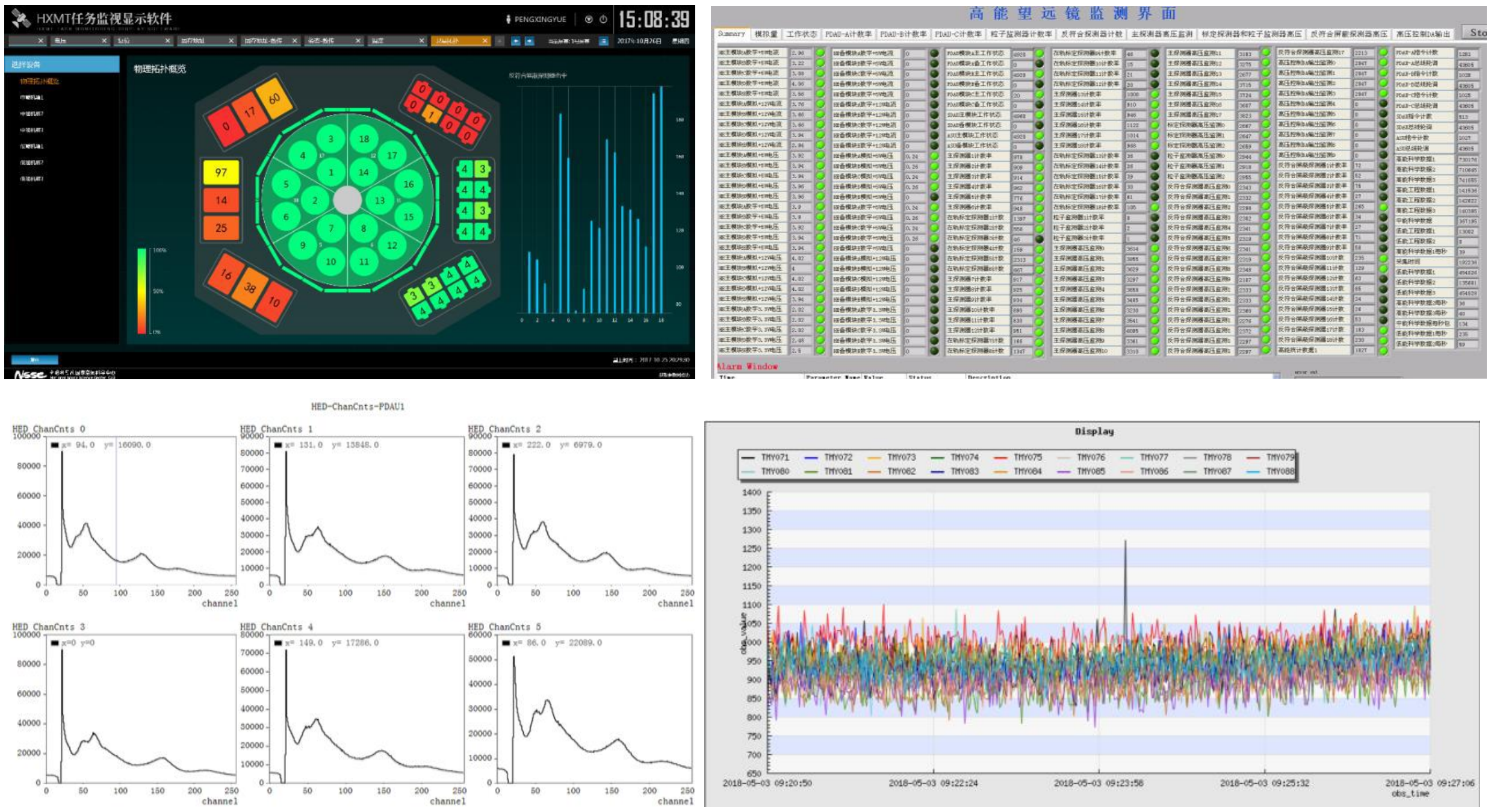

Figure 5. The payload monitoring of Insight-HXMT.

## 4.2 Payload parameters updating

In the payload monitoring, some anomalies could be found. For example, we found in ME detector, shown as Figure 6 (Left), the count rate of one pixel was too high. So we connected with ME team and investigated the reasons together, and a solutions for this problem was determined. Then we contacted with the payload PI to confirm this solution, and scheduled a parameter updating plan. Finally, the parameter updating plan was sent to MOC and uplinked to the satellite, and the anomaly was resolved. The workflow for the anomaly tracking of Insight-HXMT is shown as Figure 6 (Right).

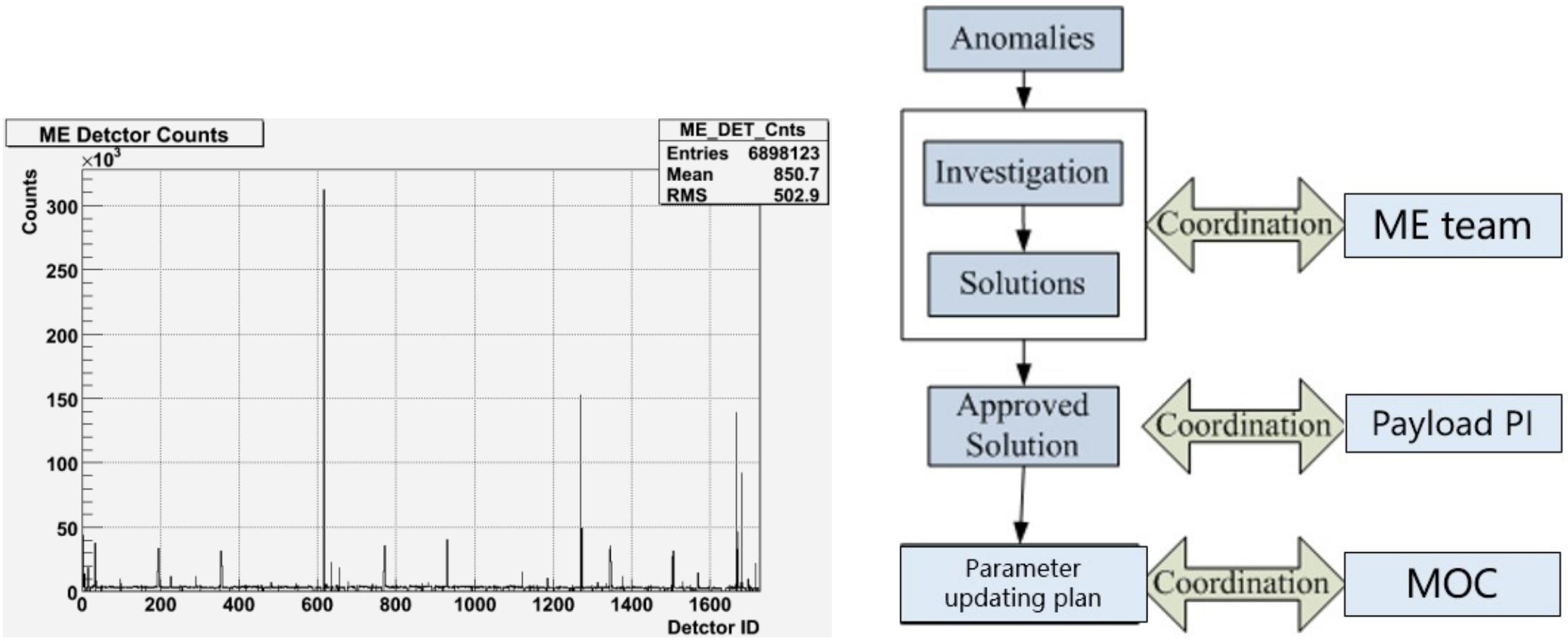

Figure 6. Anomaly tracking and on-board parameter updating.

# 5. SUMMARY

Till now, Insight-HXMT has been working for almost 1 year. All the instruments work well, and the performances and the on-orbit calibration of the instruments have been preliminary confirmed, and also some scientific results have been published [4]. Insight-HXMT has completed the whole Galactic Plane scan, and carried out the observations for pulsars, pulsar remnants, BH binaries, NS binaries and some extra-galactic sources (e.g. cluster of galaxies, BL lac).

## ACKNOWLEDGEMENTS

This work made use of the data from the Insight-HXMT mission, a project funded by China National Space Administration (CNSA) and the Chinese Academy of Sciences (CAS). The authors thank support from the National Program on Key Research and Development Project (Grant No. 2016YFA0400801) and the Strategic Priority Research Program on Space Science of the Chinese Academy of Sciences (XDA15051400).